\documentclass{PoS}

\usepackage{url}

\usepackage{natbib}
\title{UVOT Measurements of Dust and Star Formation in the SMC and M33}

\ShortTitle{UVOT Measurements of Dust and Star Formation in the SMC and M33}

\author{Lea M.\ Z.\ Hagen \\
        The Pennsylvania State University\\
        E-mail: \email{lea.zernow.hagen@gmail.com}}

\author{Michael Siegel\\
        The Pennsylvania State University\\
        E-mail: \email{siegel@swift.psu.edu}}

\author{Caryl Gronwall\\
        The Pennsylvania State University\\
        E-mail: \email{caryl@astro.psu.edu}}

\author{Erik Hoversten\\
        University of North Carolina\\
        E-mail: \email{ehoverst@live.unc.edu}}

\author{Angelica Vargas\\
        The Pennsylvania State University\\
        E-mail: \email{avargas@swift.psu.edu}}

\author{Stefan Immler\\
        NASA HQ\\
        E-mail: \email{stefan.m.immler@nasa.gov}}

\abstract{
When measuring star formation rates using ultraviolet light, correcting for dust extinction is a critical step.  However, with the variety of dust extinction curves to choose from, the extinction correction is quite uncertain.  Here, we use Swift/UVOT to measure the extinction curve for star-forming regions in the SMC and M33.  We find that both the slope of the curve and the strength of the 2175~\AA\ bump vary across both galaxies.  In addition, as part of our modeling, we derive a detailed recent star formation history for each galaxy.
}

\FullConference{Swift: 10 Years of Discovery\\
                 2-5 December 2014\\
                 La Sapienza University, Rome, Italy}

\begin{document}

% copied over from aastex
\newcommand\aj{AJ}
          % Astronomical Journal
\newcommand\araa{ARA\&A}
          % Annual Review of Astron and Astrophys
\newcommand\apj{ApJ}
          % Astrophysical Journal
\newcommand\apjl{ApJ}
          % Astrophysical Journal, Letters
\newcommand\apjs{ApJS}
          % Astrophysical Journal, Supplement
\newcommand\aap{A\&A}
          % Astronomy and Astrophysics
\newcommand\aaps{A\&AS}
          % Astronomy and Astrophysics, Supplement
\newcommand\mnras{MNRAS}
          % Monthly Notices of the RAS
\newcommand\pasp{PASP}
          % Publications of the ASP
\newcommand\nat{Nature}
          % Nature
\newcommand\ssr{Space~Sci.~Rev.}
          % Space Science Reviews

% ==========================================

\section{Introduction} \label{sec-intro}

Finding the rate at which galaxies form stars is fundamental to understanding the galaxies' formation and evolution.  One can measure the star formation rate (SFR) at many wavelengths, using both broadband and line-based methods \citep[see, e.g.,][]{kennicutt12}.  Each of these, of course, have their strengths and weaknesses.

\begin{figure}[h!]
	\centering
	\includegraphics[trim = 5mm 105mm 15mm 35mm, clip=true, width=0.6\textwidth]{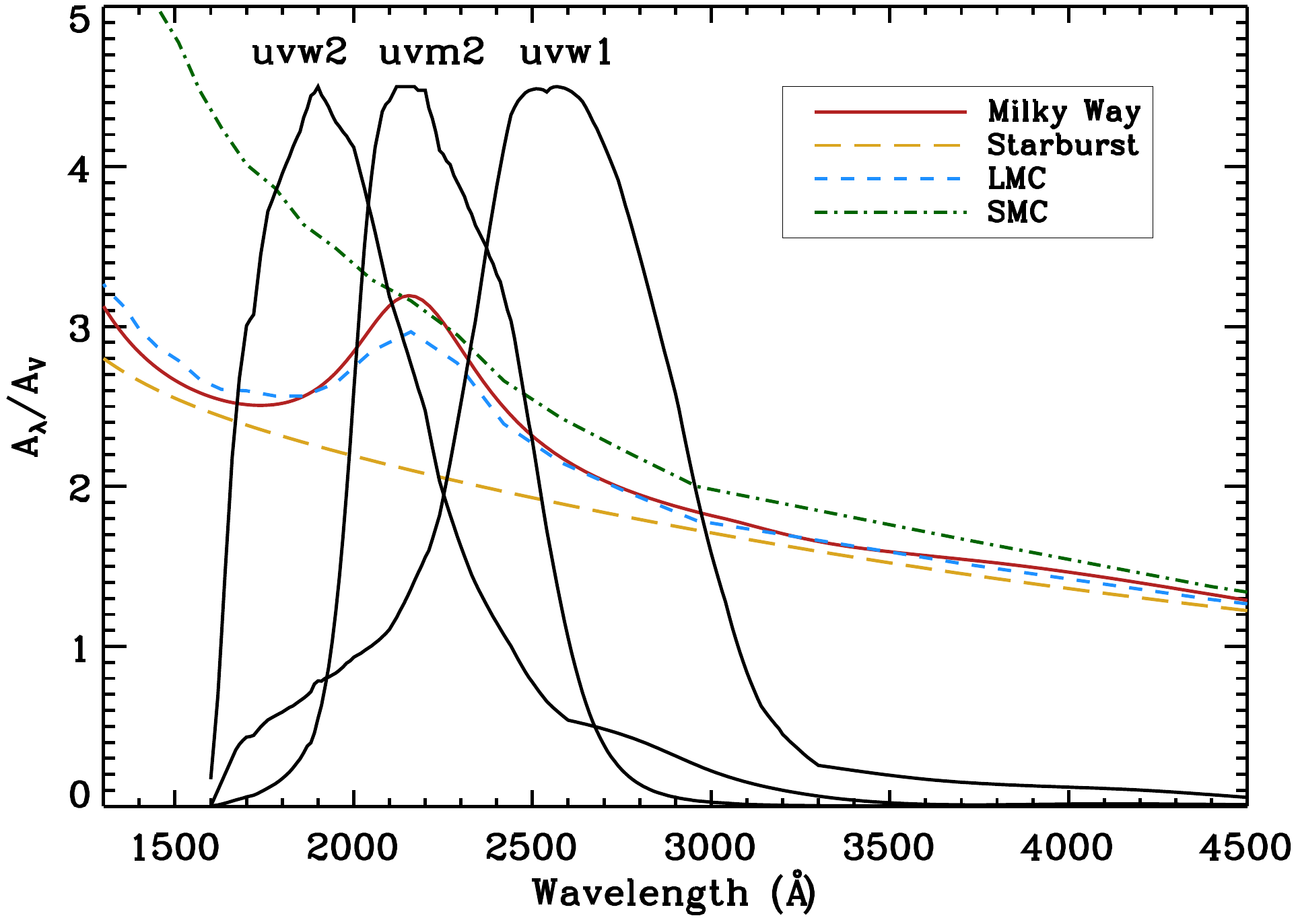}
%	trim = left bottom right top
	\caption{Several common dust extinction curves: Milky Way \citep{cardelli89}, starburst \citep{calzetti00}, Large Magellanic Cloud \citep[LMC;][]{misselt99}, and Small Magellanic Cloud \citep[SMC;][]{gordon03}.  Note the extinction curves' differing slopes and the presence (or lack) of the 2175~\AA\ dust bump.  Overlaid are the filter transmission curves for the uvw2, uvm2, and uvw1 filters, showing how their placement is useful for constraining the properties of the dust extinction curve.}
	\label{fig-dust}
\end{figure}

Ultraviolet (UV) light traces the photospheres of O and B stars, and measures star formation on timescales of $
\sim$100~Myr.  However, it is more heavily extinguished by dust than longer wavelength indicators, so it is necessary to do an accurate dust correction.  Unfortunately, the shape of the dust extinction curve is uncertain in UV.  Figure~\ref{fig-dust} shows four commonly used extinction curves, which are nearly identical at optical wavelengths, but vary widely in the UV.  There are two main differences in the extinction curves: the slope (typically parametrized as $R_V$) and the strength of the bump at 2175~\AA.

% ----------------------------
% ----------------------------

Previous measurements of the extinction curve have relied on the time-consuming procedure of observing the spectrum of one extinguished star and comparing it the spectrum of an unextinguished star of identical spectral type.  It is unsurprising, therefore, that very few extinction curve measurements have been done in this way.  The standard curve associated with the Small Magellanic Cloud (SMC), for example, is based on five lines of sight \citep{gordon03}, and the other curves rely on similarly few measurements.

\begin{figure}
	\centering
	\includegraphics[trim = 25mm 55mm 15mm 60mm, clip=true, width=0.45\textwidth]{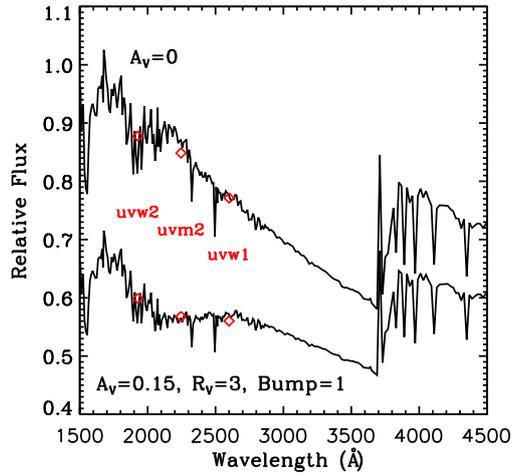}
%	trim = left bottom right top
	\caption{Model spectra representing the ability of UVOT to measure dust extinction properties.  The top spectrum has no dust attenuation ($A_V = 0$).  The bottom spectrum has a small attenuation of $A_V = 0.15$ with a dust extinction curve similar to that of the Milky Way.  The UVOT fluxes for each spectrum are marked with red diamonds.  Compared to the top spectrum, the uvm2 flux is suppressed relative to uvw1 and uvw2 (which traces the 2175~\AA\ bump), and the uvw1 and uvw2 fluxes are more similar (which traces $R_V$).}
	\label{fig-spec}
\end{figure}

Our goal is to model the dust extinction curves in the SMC and M33 and determine if the curves vary across the galaxies.  The UV/optical telescope \citep[UVOT;][]{roming05} on Swift \citep{gehrels04} has three near-UV filters, which are included in Figure~\ref{fig-dust}.  The uvm2 filter overlaps the 2175~\AA\ dust bump, and the uvw1 and uvw2 filters are to either side, which means we can constrain the dust curve.  This is demonstrated further in Figure~\ref{fig-spec}: comparing the unextinguished spectrum to the extinguished one, the uvm2 magnitude is suppressed relative to uvw1 and uvw2, and a line connecting uvw1 and uvw2 becomes shallower.  The degree to which these happen corresponds to the strength of the 2175~\AA\ bump and the value of $R_V$, respectively.

% ----------------------------
% ----------------------------

To accomplish this goal, we use a mosaic of UVOT and ground-based optical observations of the SMC and M33, the details of which are described in Section~\ref{sec-data}.  We model the spectral energy distributions of star-forming regions in the galaxies in Section~\ref{sec-model}, and describe our early results in Section~\ref{sec-res}.

% ==========================================

\section{Data} \label{sec-data}

%\subsection{UV/Optical Data}

\begin{figure}
	\centering
	\includegraphics[trim = 0mm 0mm 0mm 0mm, clip=true, width=0.4\textwidth]{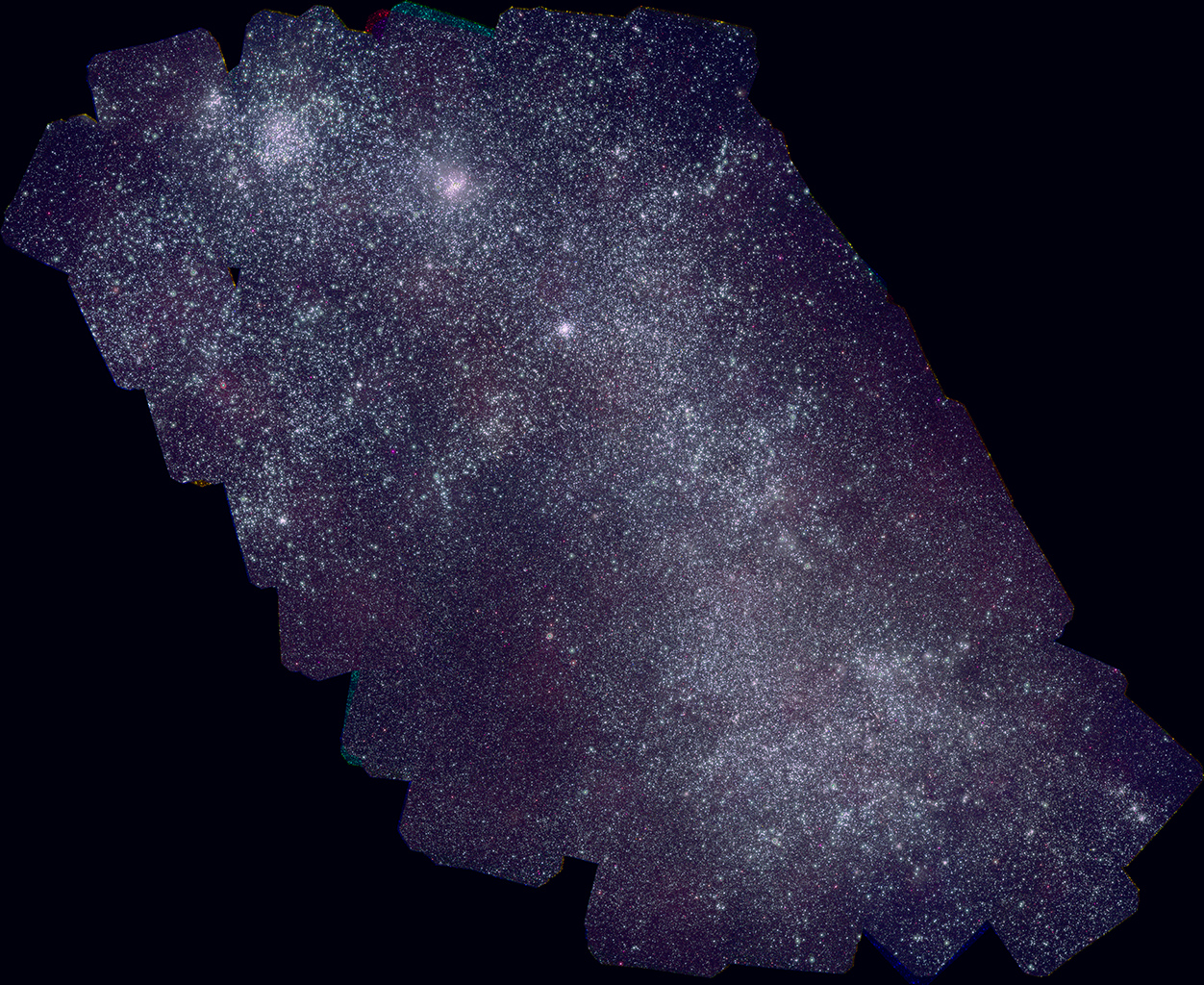}
	\ \ 
	\includegraphics[trim = 0mm 0mm 0mm 0mm, clip=true, width=0.45\textwidth]{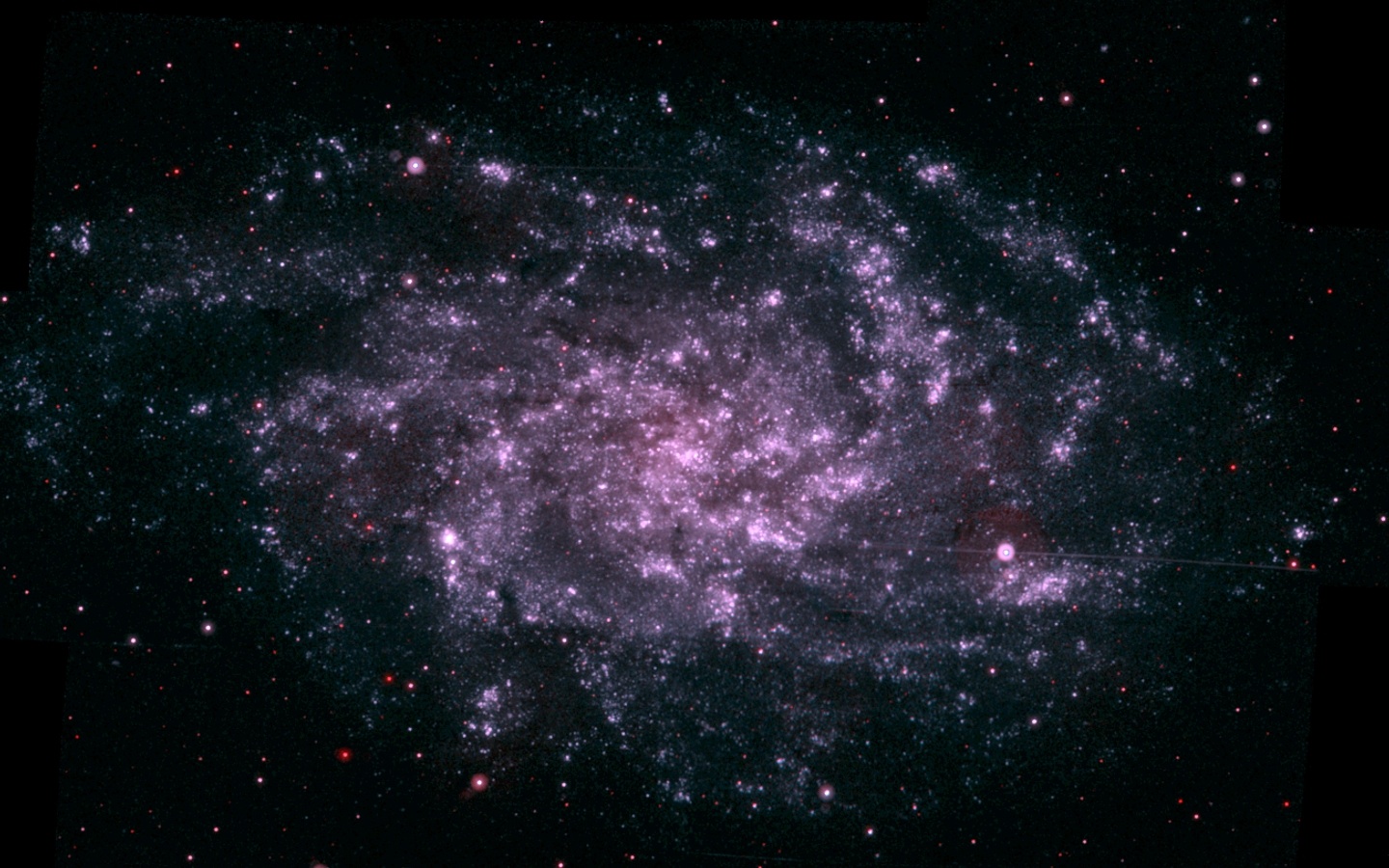}
%	trim = left bottom right top
	\caption{False color UVOT images of the SMC (left) and M33 (right).  The image of the SMC (M33) is about 2.3$^\circ$ (1.2$^\circ$) across, corresponding to 2.4~kpc (17.6~kpc) at the assumed distance of 60~kpc (840~kpc).}
	\label{fig-color}
\end{figure}

We observed the SMC with UVOT as part of the SUMaC (Swift Ultraviolet survey of the Magellanic Clouds) program.  The SMC observations consist of 50 individual tiles, each $17 \times 17$ arcminutes, and a total exposure time of 1.8~days.
The M33 observing campaign consisted of 11 hours of exposure divided between 13 tiles.
False-color mosaics of the SMC and M33 UV imaging are in Figure~\ref{fig-color}. 
The UVOT data reduction followed the procedure discussed in \citet{siegel14}.

% ----------------------------
% ----------------------------

%The UVOT data reduction followed standard procedures, and
%is described in the UVOT Software Guide.\footnote{http://heasarc.gsfc.nasa.gov/docs/swift/analysis}
%Exposure maps and images were generated with UVOT FTOOLS (HEAsoft 6.6.1).\footnote{http://heasarc.gsfc.nasa.gov/docs/software/lheasoft/}
%This involves two flux conserving interpolations of the images;
%the first of these converts from the raw frame to sky coordinates, and the second occurs when summing the images.  During processing, a correction is applied for known bad pixels.

We used archival ground-based optical imaging of both the SMC and M33.  The SMC optical data were from \citet{massey02}.  The data were taken at the Curtis Schmidt telescope at CTIO in the Harris \textit{UBVR} filters \citep{massey00}.  Due to difficulties calibrating the \textit{U}-band data, we discarded it from analysis.
The M33 data were from \citet{massey06}, taken at the Mayall telescope at KPNO in the Harris \textit{UBVRI} filters.

%\subsection{Data Reduction}

All of the UVOT and optical images were aligned and rebinned to $10''$ (2.9~pc) pixels for the SMC, and to $2.45''$ (10~pc) for M33, using SWarp \citep[version 2.19.1;][]{bertin02}.
To remove the diffuse background in each image, we used a circular median filtering technique following \citet{hoversten11}.  \citet{pleuss00} used HST imaging of HII regions in M101 to show that they range between 20~pc and 220~pc in size; we therefore adopt this range of values for the circles' diameters.

Finally, we used Source Extractor \citep[SE; version 2.5.0;][]{bertin96} to identify star-forming regions in the UVOT uvw2 image.  We chose this filter because it is the bluest available, thus tracing the most massive young stars.  (The uvw2 filter has a red leak, but the transmission isn't significant beyond $\sim$3000~\AA.)  SE has a known problem in which it can identify pixels in non-contiguous regions as belonging to the same region; to correct for this, we follow the method described in Appendix~A of \citet{hoversten11}.
After this correction, we find 396 star-forming regions in the SMC and 656 in M33.  We then use the SE-defined regions to extract photometry in the other UV and optical filters.

% ==========================================

\section{Modeling} \label{sec-model}

% explain pegase 2.0 grids, fitting using colors, masses from normalization

We model the spectral energy distributions of each star-forming region by comparing our data to a grid of models.  We create these models using the PEGASE.2 spectral synthesis code \citep{fioc97}.  The grid includes spectra for ages of 1~Myr to 13~Gyr, total dust attenuation ($A_V$) from 0 to 7 magnitudes, dust extinction curve slopes ($R_V$) from 1.5 to 5.0, and 2175~\AA\ bump strength from 0 to 2 (where 1 is the strength found in the Milky Way).
We assume that since we have isolated individual star-forming regions, a single starburst is a reasonable star formation history.

At each point in the grid, we compared the colors of each star-forming region to those of the model and calculated a chi-squared.  By comparing colors, and not magnitudes, we could account for the mass normalization at a later stage.  We repeated this 250 times while moving the magnitudes within their uncertainties.  We then created a distribution of the best fits from each of the 250 runs, from which we extracted an overall best fit and uncertainties.  The mass normalization was derived based on this best fit.

In upcoming papers, this chi-squared procedure will be replaced with a more statistically robust Markov Chain Monte Carlo fitting technique.  This will enable better measurements of degeneracies and more reliable uncertainties.  We will also fit the magnitudes, rather than the colors, and include the mass as a free parameter.

% ==========================================

\section{Early Results} \label{sec-res}

\begin{figure}
	\centering
	\includegraphics[trim = 35mm 45mm 20mm 75mm, clip=true, width=0.3\textwidth]{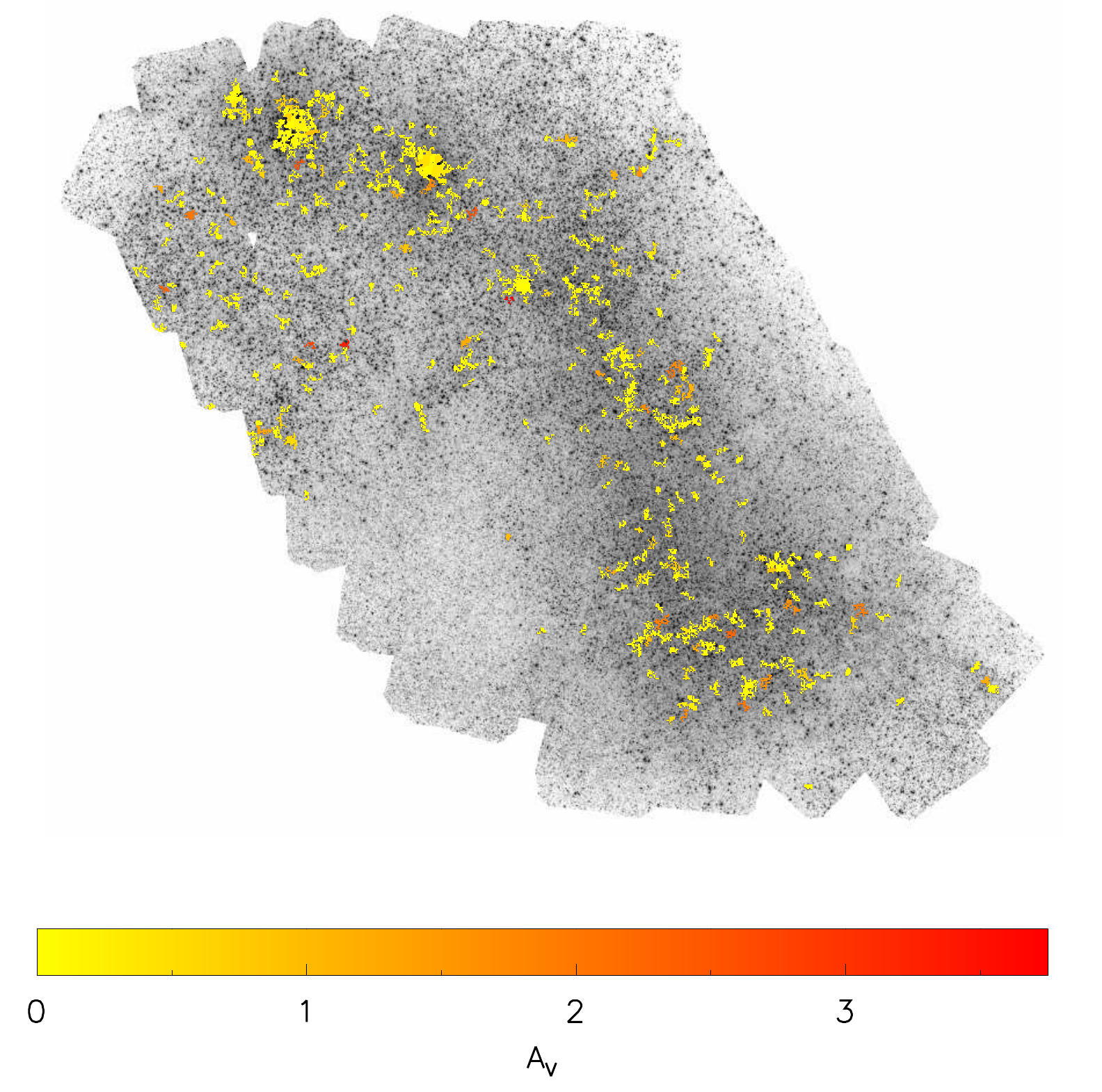}
	\
	\includegraphics[trim = 35mm 45mm 20mm 75mm, clip=true, width=0.3\textwidth]{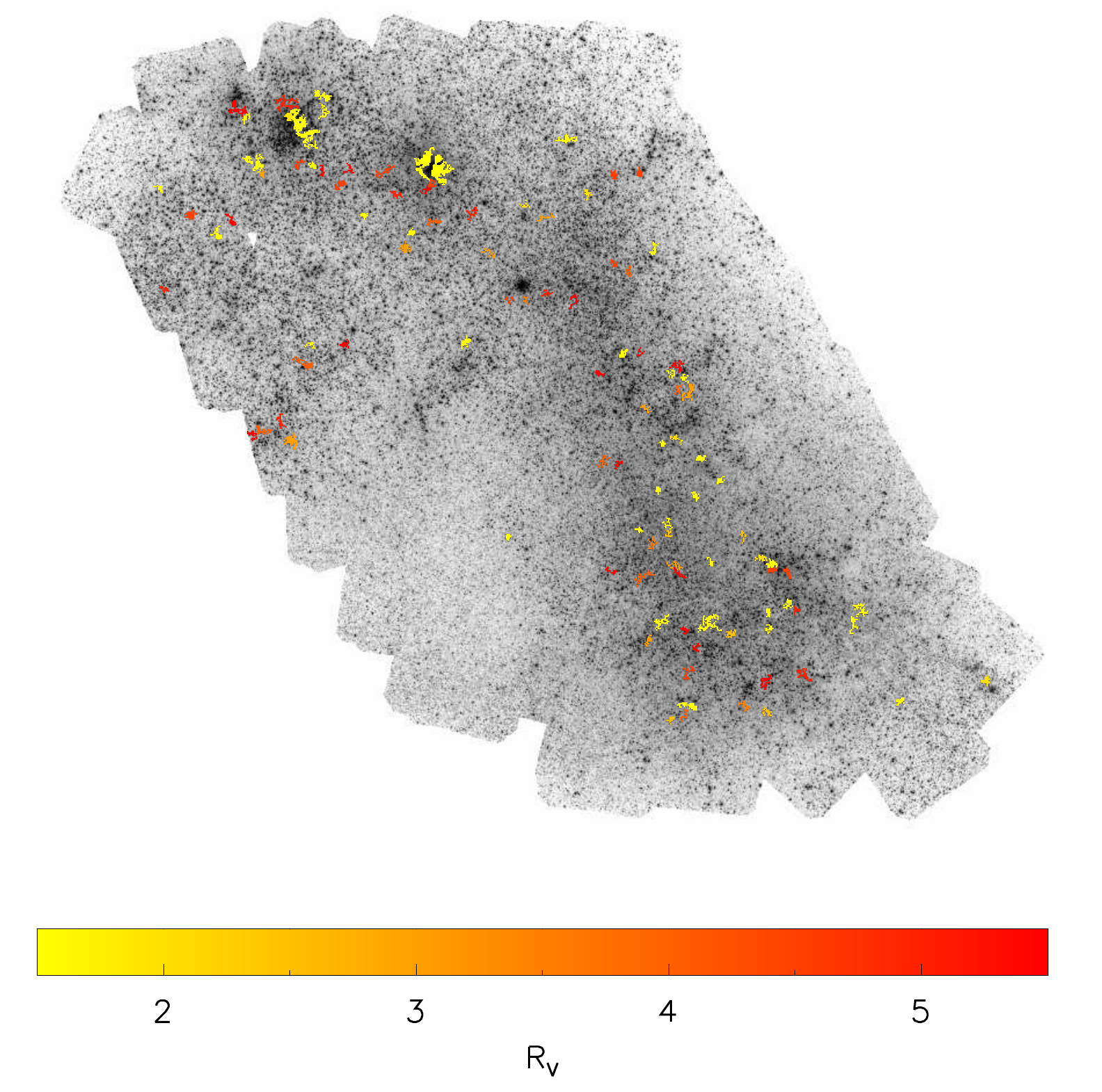}
	\
	\includegraphics[trim = 35mm 45mm 20mm 75mm, clip=true, width=0.3\textwidth]{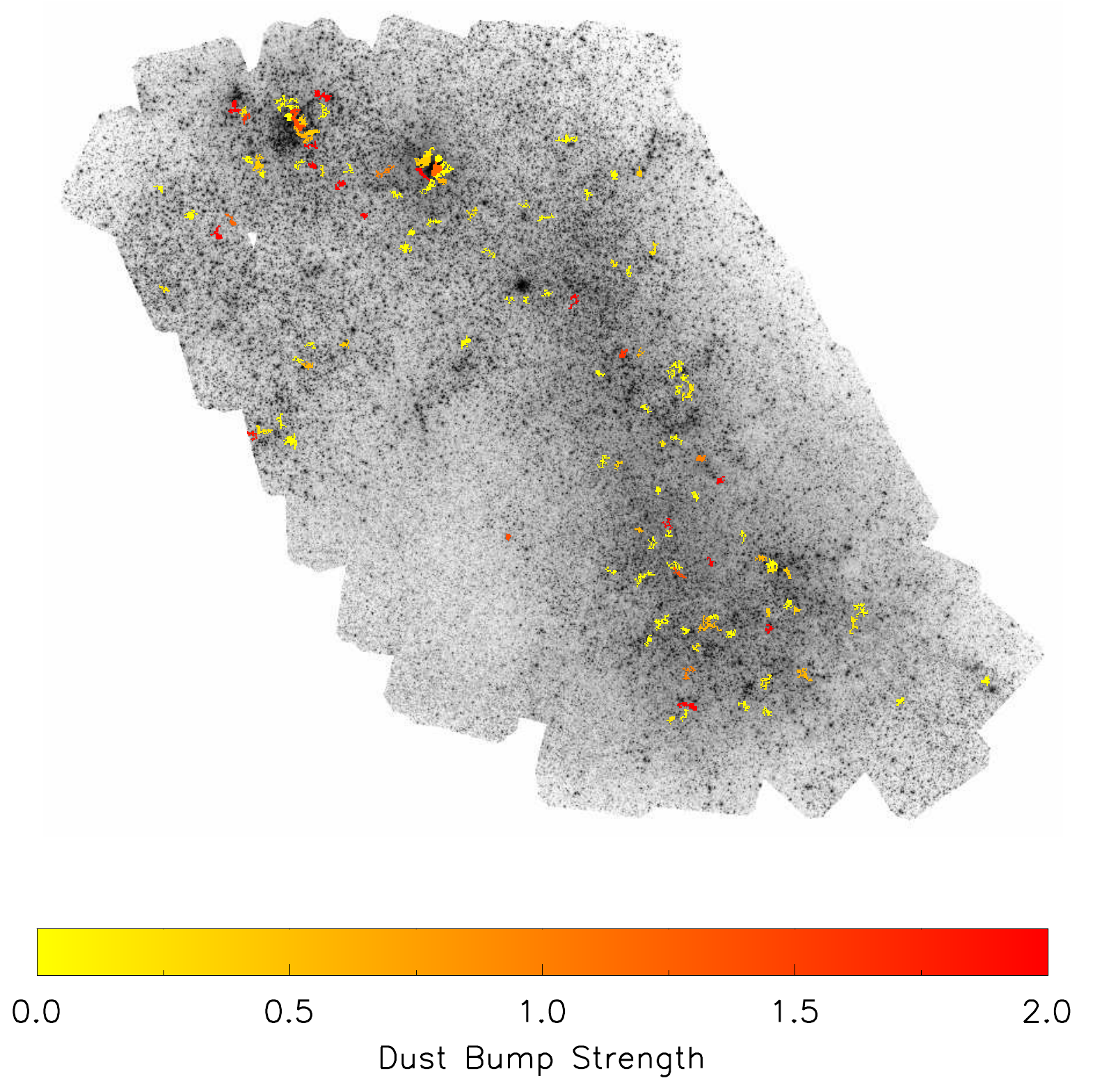}
	
	\includegraphics[trim = 35mm 45mm 20mm 75mm, clip=true, width=0.3\textwidth]{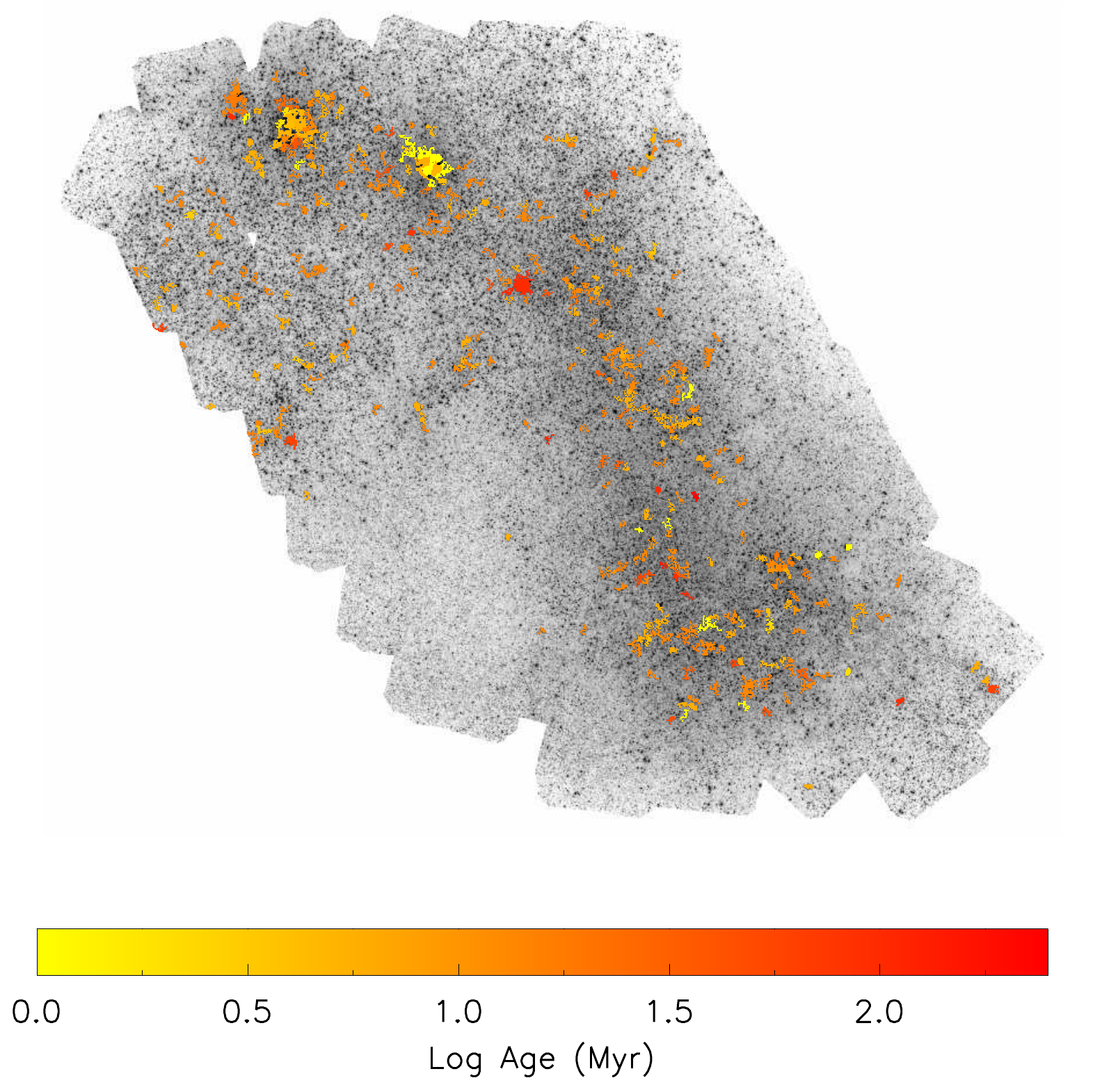}
	\ \ \ \
	\includegraphics[trim = 20mm 55mm 20mm 55mm, clip=true, width=0.33\textwidth]{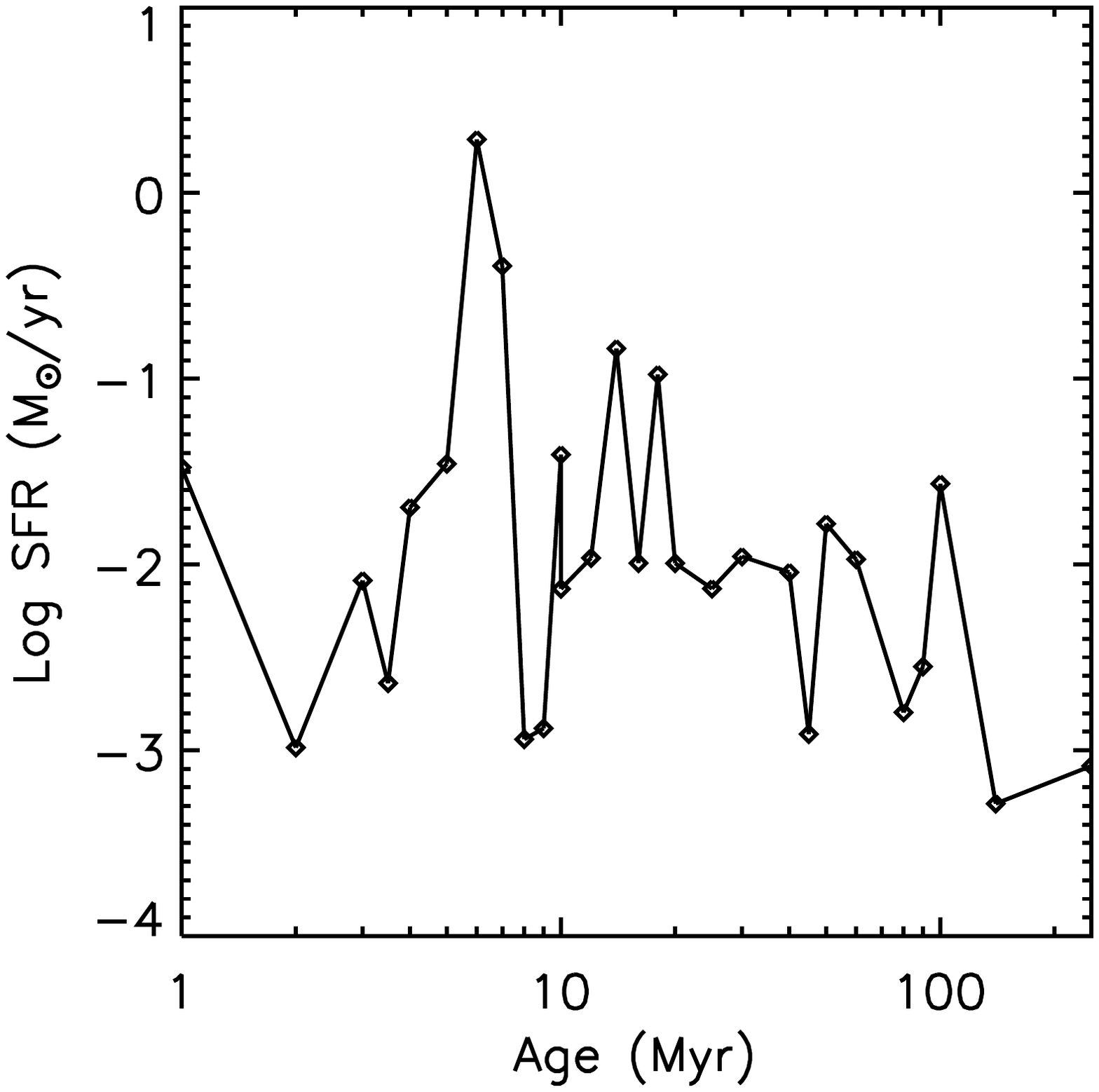}
%	trim = left bottom right top
	\caption{\textit{Top row:} Maps of dust properties for star-forming regions in the SMC.  We assume that we don't constrain $R_V$ or the bump strength in regions where $A_V$ is less than 0.1.
	\textit{Bottom row:} Map of the ages of star-forming regions, and the corresponding star formation history.}
	\label{fig-res_smc}
\end{figure}

\begin{figure}
	\centering
	\includegraphics[trim = 35mm 45mm 20mm 60mm, clip=true, width=0.3\textwidth]{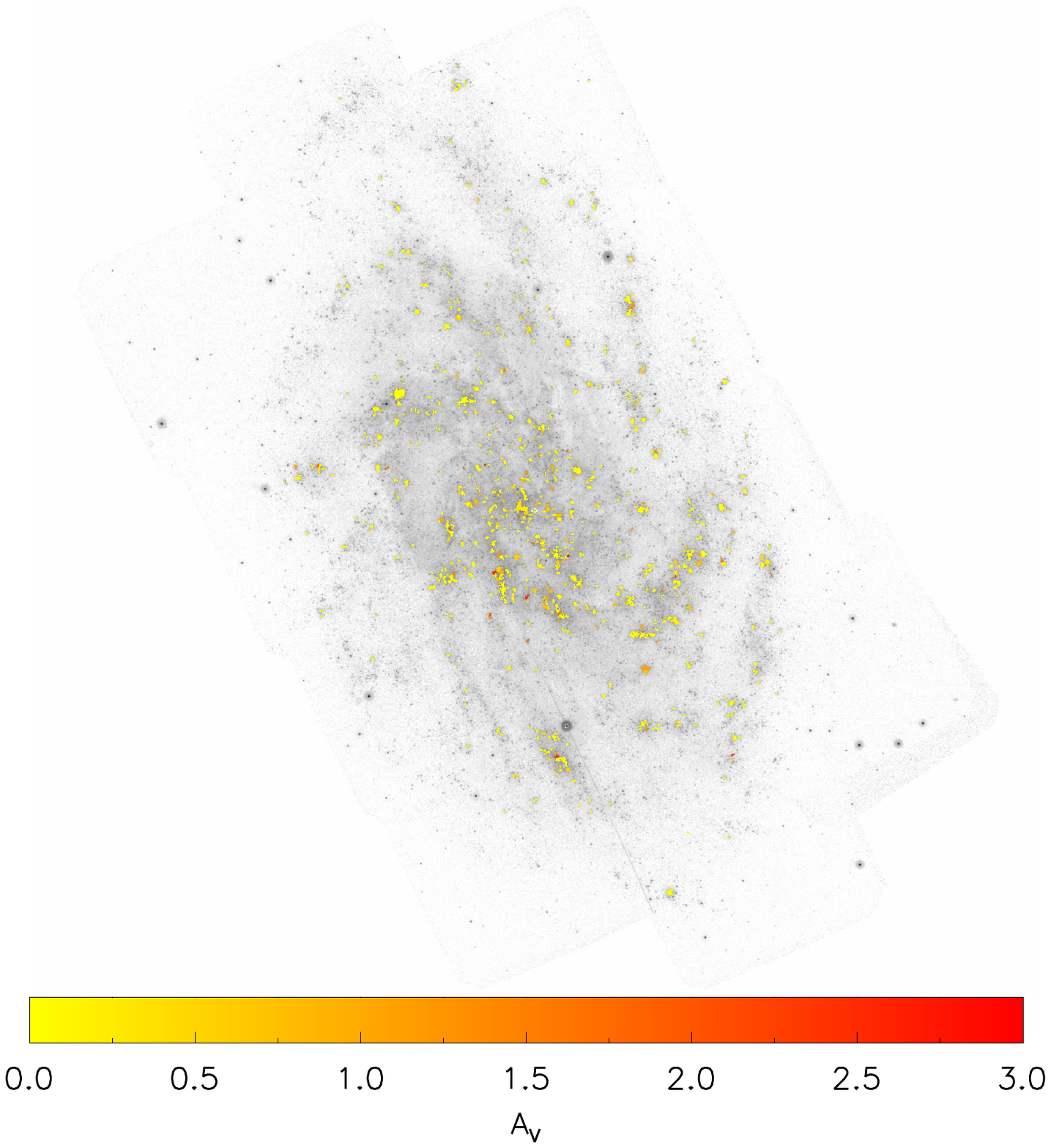}
	\
	\includegraphics[trim = 35mm 45mm 20mm 60mm, clip=true, width=0.3\textwidth]{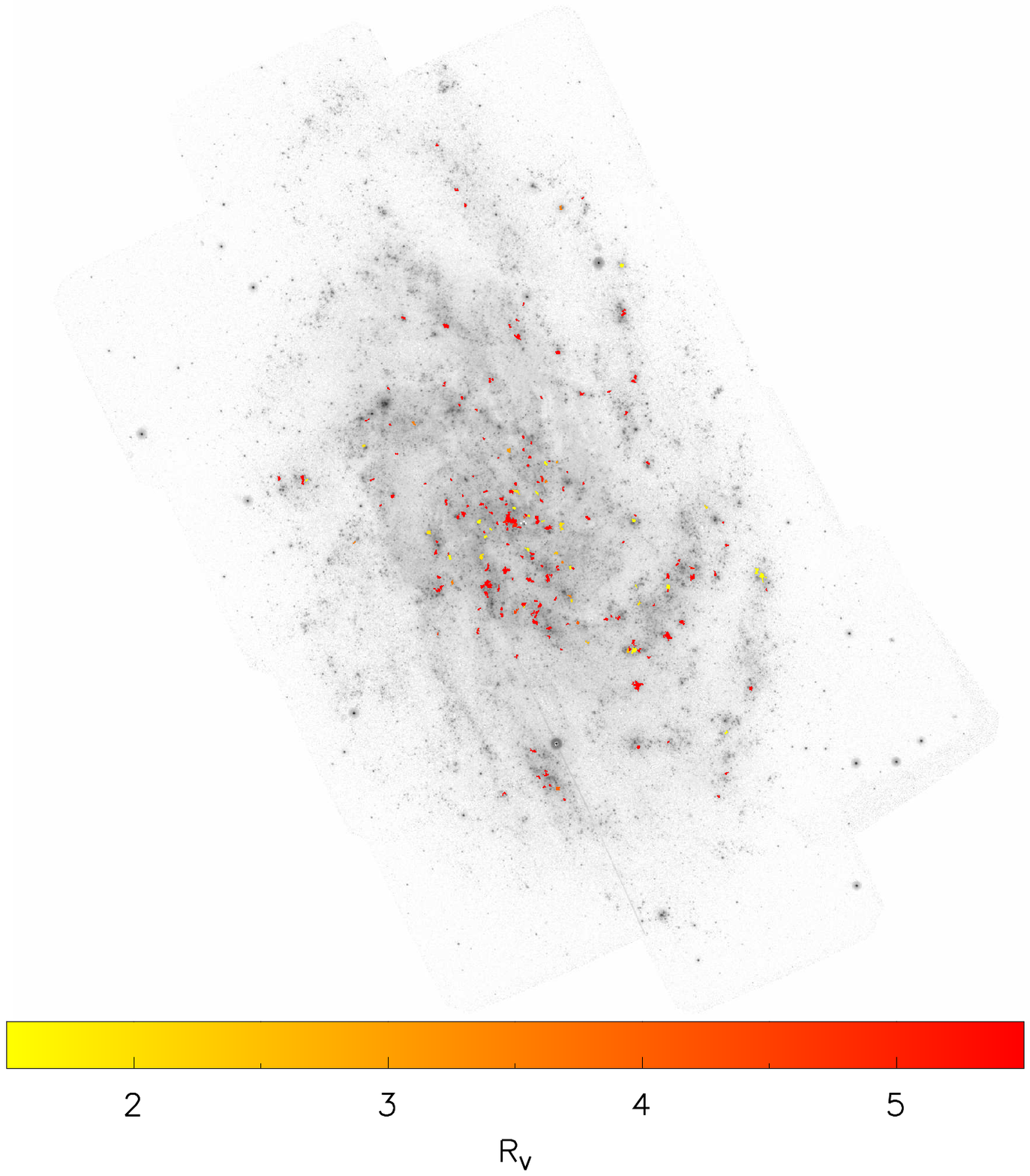}
	\
	\includegraphics[trim = 35mm 45mm 20mm 60mm, clip=true, width=0.3\textwidth]{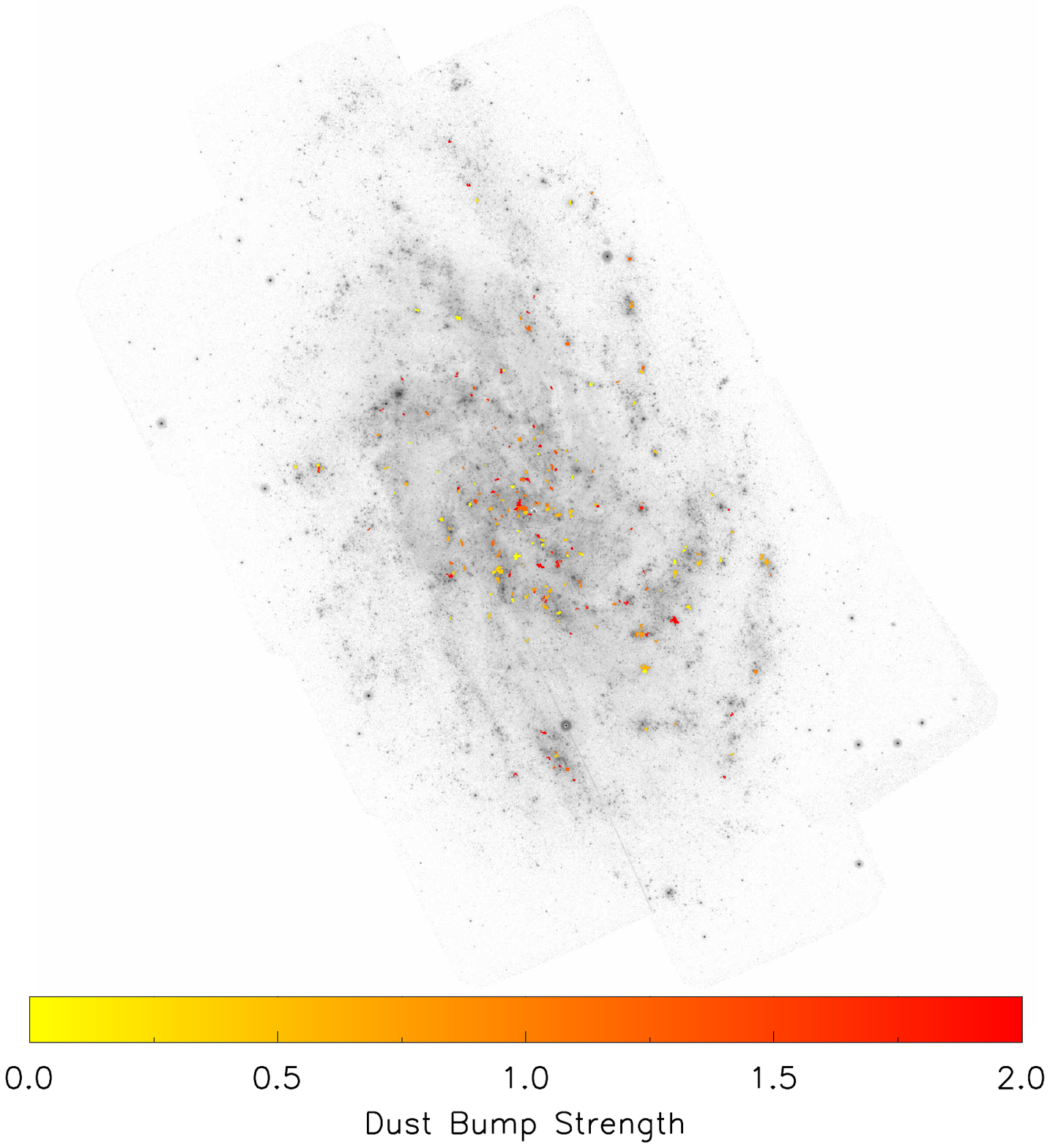}
	
	\includegraphics[trim = 35mm 45mm 20mm 60mm, clip=true, width=0.3\textwidth]{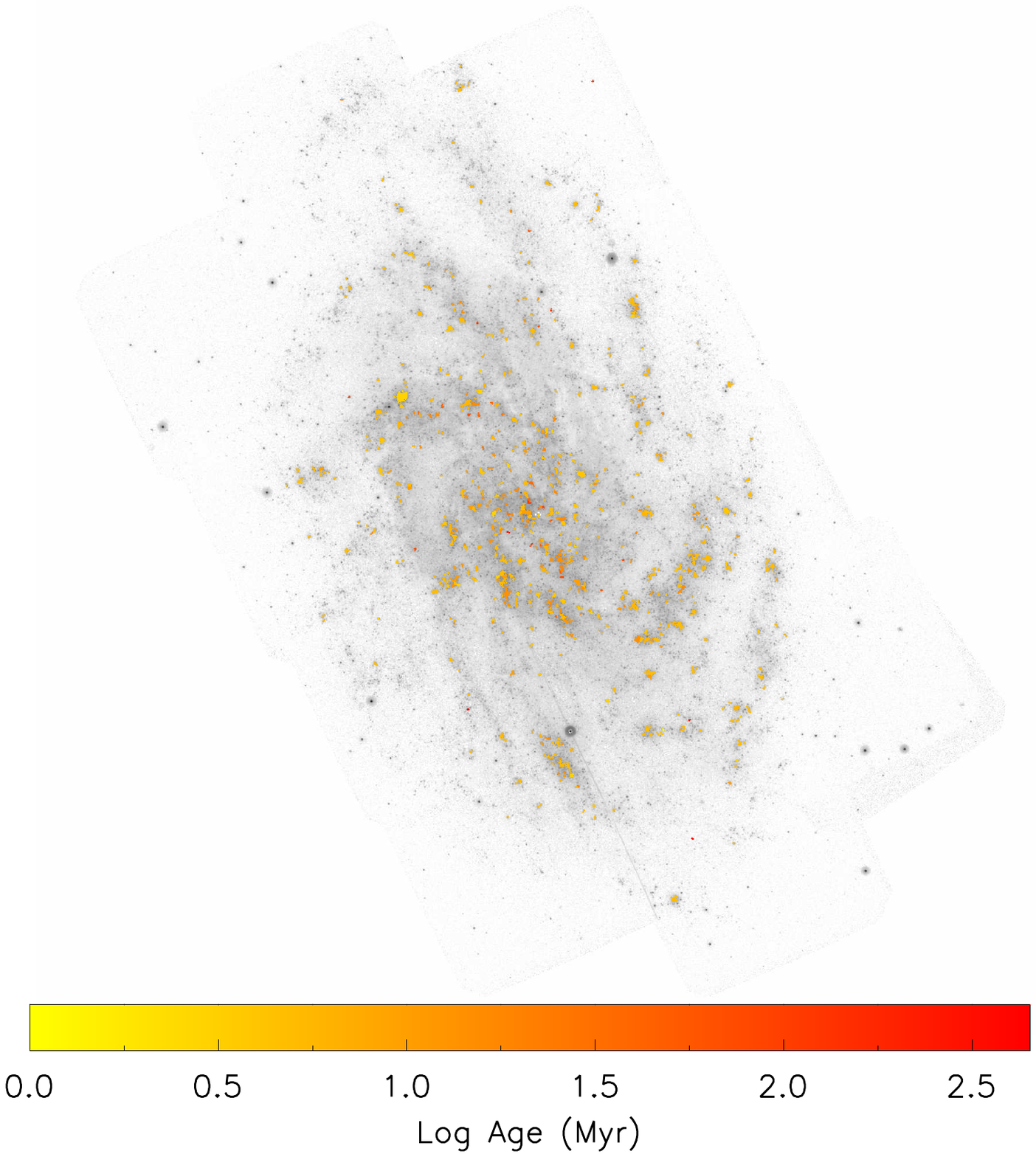}
	\ \ \ \
	\includegraphics[trim = 20mm 55mm 20mm 55mm, clip=true, width=0.33\textwidth]{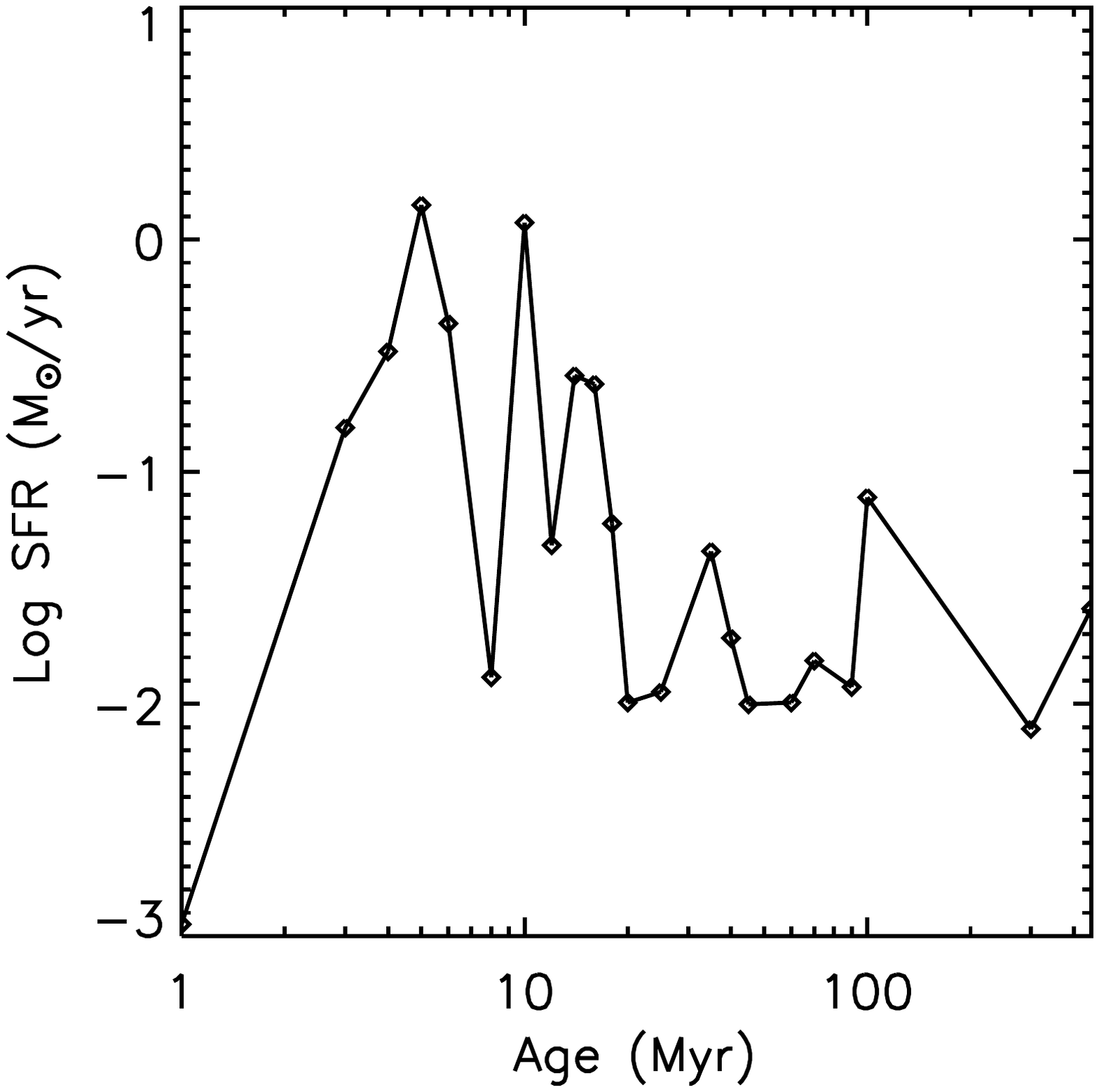}
%	trim = left bottom right top
	\caption{Same as Figure~\protect\ref{fig-res_smc}, but for M33.}
	\label{fig-res_m33}
\end{figure}

The best-fit properties of the star-forming regions in the SMC and M33 are mapped in Figures~\ref{fig-res_smc} and~\ref{fig-res_m33}.  Also in these figures are the star formation histories derived from the regions' masses (not shown) and ages.

The most important result is that the dust extinction curve varies significantly across both the SMC and M33.  For studies of objects within these galaxies, the best-fit extinction curve can be used in place of others.  In particular, the SMC curve (plotted in Figure~\ref{fig-dust}) does not appear to apply to young star-forming regions in the SMC.
For most other dust corrections in the literature, by necessity, authors choose their favorite extinction curve with little ado.  This work shows how variable the extinction curve is, and therefore can be used to quantify the systematic error when correcting for dust extinction.

The second important result is the detailed star formation history of the star-forming regions over the past few hundred million years.  When using optical data, it is difficult to distinguish between the most massive stars, because it's already in the Rayleigh-Jeans tail.  As such, measurements of the youngest ages are difficult.  By utilizing ultraviolet light, we can better resolve the ages of the youngest populations, and better understand the most recent star formation.

In upcoming work, we will also be mapping the dust properties and ages across the entirety of both galaxies.  Although the resolution will be coarser (due to computational constraints), this will yield useful information for more than just the brightest regions of star formation.

% ----------------------------
% ----------------------------

% ----------------------------
% ----------------------------

%\begin{thebibliography}{99}
%  \bibitem{...} ....
%\end{thebibliography}

\bibliographystyle{apj}

% http://tex.stackexchange.com/questions/163559/spacing-between-lines-in-per-entry-of-bibliography
\begingroup
	\setlength{\bibsep}{2pt}
	\linespread{1}\selectfont
	\bibliography{master_paper_list}
\endgroup

\end{document}